# Characterizing Variability in Bright Metallic-Line A (Am) Stars Using Data from the NASA TESS Spacecraft

*Joyce Ann Guzik*
*Los Alamos National Laboratory, MS T-082, Los Alamos, NM 87545*
*joy@lanl.gov*

*Jason Jackiewicz*
*Department of Astronomy, New Mexico State University, Las Cruces, NM*

*Giovanni Catanzaro*
*INAF--Osservatorio Astrofisico di Catania, Via S. Sofia 78, I-95123 Catania, Italy*

*Michael S. Soukup*
*Los Alamos National Laboratory (retired), Albuquerque, NM 87111*

**Abstract**

Metallic-line A (Am) stars are main-sequence stars of around twice the mass of the Sun that show element abundance peculiarities in their spectra. The radiative levitation and diffusive settling processes responsible for these abundance anomalies should also deplete helium from the region of the envelope that drives $\delta$ Scuti pulsations. Therefore, these stars are not expected to be $\delta$ Scuti stars, which pulsate in multiple radial and nonradial modes with periods of around 2 hours. As part of the NASA TESS Guest Investigator Program, we proposed photometric observations in 2-minute cadence for samples of bright (visual magnitudes around 7-8) Am stars. Our 2020 SAS meeting paper reported on observations of 21 stars, finding one $\delta$ Scuti star and two $\delta$ Scuti / $\gamma$ Doradus hybrid candidates, as well as many stars with photometric variability possibly caused by rotation and starspots. Here we present an update including 34 additional stars observed up to February 2021, among them three $\delta$ Sct stars and two $\delta$ Sct / $\gamma$ Dor hybrid candidates. Confirming the pulsations in these stars requires further data analysis and follow-up observations, because of possible background stars or contamination in the TESS CCD pixels with scale 21 arc sec per pixel. Asteroseismic modeling of these stars will be important to understand the reasons for their pulsations.

## 1. Introduction

Stars of nearly every type and evolutionary state show pulsational variability (Aerts, Christensen-Dalsgaard, and Kurtz 2010). The pulsation properties can be used to infer the internal structure and processes in stars, and validate stellar model physics and theoretical interpretations. This paper focuses on two types of main-sequence (core hydrogen-burning) variable stars that pulsate in one or more radial and nonradial modes, namely the $\gamma$ Doradus and $\delta$ Scuti variables. Greek letters such $\gamma$ = gamma, and $\delta$ = delta are used to refer to the brightest stars in a constellation in order of the Greek alphabet, and many variable star types are named after their prototype star.

Here we discuss results of searches for pulsating variables using photometric data obtained by the Transiting Exoplanet Survey Satellite (TESS) as part of their Guest Investigator Program. This paper updates results presented at the 2020 SAS meeting and documented in the Proceedings.

The $\delta$ Sct variables are main-sequence (core hydrogen-burning) or slightly post-main-sequence (shell hydrogen-burning) variables about twice the mass of the Sun, which pulsate in multiple radial and nonradial modes with periods of about 2 hours, or frequencies about 12 cycles/day (Aerts et al. 2010, Breger 2000). The cause of their pulsations is believed to be the 'kappa', or opacity valving mechanism, named for the Greek letter $\kappa$ (kappa) used to represent opacity. Opacity is a measure of how 'opaque', i.e., resistant, the stellar interior plasma is to transfer of photon radiation through the layer. Figure 1 shows the opacity versus interior temperature from a two solar-mass model. For the $\delta$ Sct stars, the increased opacity region labeled He+ where the second electron of helium is ionizing at 50,000 K in the stellar envelope is at an optimum location to block radiation, causing this layer to absorb heat, expand, and then cool and contract in a feedback loop to produce the pulsations.

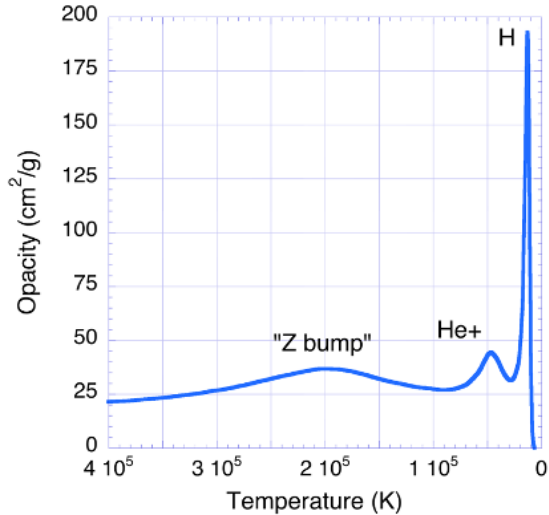

Figure 1: Opacity vs. interior temperature for 2 solar-mass main-sequence stellar model. In this plot, the stellar interior corresponds to higher temperatures toward the left, while the stellar surface is toward the right. The opacity enhancement at around 50,000 K (labeled He+) where the second electron of helium is being ionized is responsible for driving δ Sct pulsations. The "Z bump" is responsible for pulsations in more massive stars.

The γ Dor variables are about 1.5 times the mass of the Sun and pulsate in nonradial gravity modes with periods of 1 to 3 days (Kaye et al. 1999). The convective blocking mechanism at the base of the envelope convection zone (Guzik et al. 2000) is proposed to drive the pulsations. The pulsation driving arises because the pulsation period is shorter than the local convective timescale at the base of the envelope convection zone, and so the convection cannot adapt quickly enough during a pulsation cycle to transport the radiation emerging from the stellar center, causing radiation flow to be blocked periodically.

Before the high-precision long time-series photometric observations made by the *Kepler* spacecraft (Borucki et al. 2010; Gilliland et al. 2010), theory and stellar models for the most part explained the locations of the δ Sct and γ Dor instability regions in the H-R diagram, and hybrid variables pulsating in both low- and high-frequency modes were found in the overlapping region between these two instability regions (see Fig. 12). However, the *Kepler* data showed hybrid stars scattered throughout both instability regions, and even beyond the edges (Grigahcene et al. 2010; Uytterhoeven et al. 2011).

The metallic-line A (Am) stars show significant underabundances of Ca and Sc, and enhanced abundances of Ti and Fe-group elements. For the Am stars, the diffusive processes responsible for the abundance peculiarities are also expected to cause helium to settle out of the pulsating driving region, making the δ Sct pulsation driving mechanism ineffective (Breger 1970; Kurtz 1976). Nevertheless, some Am stars do show pulsations, raising questions about whether another pulsation mechanism is responsible (see, e.g., Smalley et al. 2017 and Murphy et al. 2020). Other pulsation driving mechanisms also are being explored to explain the prevalence of hybrid stars throughout the γ Dor and δ Sct instability regions. See Guzik (2021) for a recent review of outstanding questions surrounding δ Sct and γ Dor stars.

## 2. Am Star Results from TESS Guest Investigator Program

The NASA TESS spacecraft was launched on April 18, 2018 and is in an elliptical 13.7-day lunar resonance orbit around Earth. Its main mission is to search for transiting exoplanets around nearby G, K, and M spectral-type stars using the transit method (Riker et al. 2015). The TESS spacecraft has four CCD cameras aligned in a vertical strip to view a 24° by 90° section of the sky (called a 'sector') continuously for 27 days before moving to the next (partially overlapping) sector. The sky below the ecliptic plane was observed in 13 sectors for the first year of operation (Cycle 1), and above the ecliptic plane in 13 sectors (Cycle 2). TESS has finished its main mission, and is now in extended mission, currently observing 13 sectors, again below the ecliptic plane (Cycle 3). Because the sectors overlap, targets near the north and south ecliptic poles are observable for up to a year. Data products include full-frame images with 30-minute cadence, as well as 2-min cadence light curves for several hundred thousand stars in each sector. Beginning in Cycle 3, TESS is taking 10-minute cadence full-frame images and also offers 20-sec cadence light curves on a smaller selection of targets. TESS data is available from the Mikulski Archive for Space Telescopes (MAST, https://archive.stsci.edu/). This paper includes analysis of observations through Sector 34, ending February 8, 2021.

Catanzaro et al. (2019) used high-resolution spectroscopy + Gaia DR2 parallaxes to determine stellar parameters and detailed abundances for 62 metallic-line A (Am) stars. For Cycle 2 of the TESS Guest Investigator Program, we requested two-minute cadence observations of these stars, most of which have TESS magnitudes 7-8. Because the stars in the Catanzaro sample were at declinations visible from the Northern Hemisphere, while Cycle 3 observations were targeting regions of the sky below the ecliptic plane, we decided to expand the target list using the catalog of non-magnetic chemically peculiar stars of Paunzen et al. (2013), choosing 49 stars with similar characteristics to the Catanzaro et al. sample, but observable by TESS during Cycle 3. See TESS GI programs G022027 and G03060 at https://heasarc.gsfc.nasa.gov/docs/tess/approved-programs.html for a list of observations to date.

## 3. TESS Amplitude Spectra for δ Sct and Hybrid δ Sct/γ Dor Pulsators

So far, 32 of the Am stars in the Catanzaro et al. (2019) sample (Table 1) and 23 stars of the Paunzen et al. (2013) sample (Table 2) have been observed in 2-min cadence. In Figs. 2 through 9, we show the amplitude spectra for the four δ Sct stars and four δ Sct/γ Dor hybrid candidates, calculated using Fourier transforms of the time-series data. The δ Sct stars show multiple peaks at periods >5 c/d, and the hybrid candidates show in addition one or more peaks with significant amplitude (>20 ppm) around 1 c/d. According to the SIMBAD database, none of these stars were previously known δ Sct or γ Dor variables.

Some of the Catanzaro et al. stars have 2-min cadence observations during more than one sector. Having a second 27-day sector of observations sharpens the frequency resolution by a factor of two and also increases the signal-to-noise ratio by a factor of sqrt(2). Longer time series are important to resolve the lower frequency γ Dor pulsations, as well as to resolve rotationally split frequencies (see Fig. 2b).

Our 2020 SAS proceedings paper gives examples of light curves showing other types of photometric variability. These variations may be due to rotation and magnetic activity generating starspots, or may be the result of the orbital motion of a close interacting binary. The binary interpretation is doubtful because Catanzaro et al. (2019) screened the sample to remove spectroscopic binaries. Nevertheless, an eclipsing binary, HD 188854, also known as V2094 Cyg, with an orbital period of 8.5 days, previously discovered from *Kepler* data to have a γ Dor component (Çakırlı 2015), was included in the Catanzaro et al. (2019) sample. Periodicities caused by rotation/starspots and γ Dor pulsations can be similar. In addition, low-frequency peaks may be global Rossby waves (Saio et al. 2019) instead of γ Dor g-mode pulsations. Because the cause of the low-frequency variability is not confirmed, we refer to possible γ Dor and δ Sct/γ Dor hybrid stars as 'candidates'.

In the Fig. 2-9 captions, we give the number of frequencies in the amplitude spectrum having signal-to-noise ratio > 4. These frequencies are calculated by successively pre-whitening the light curve to subtract out the signal from the highest amplitude peak. Some of these frequencies are combinations of higher amplitude frequencies.

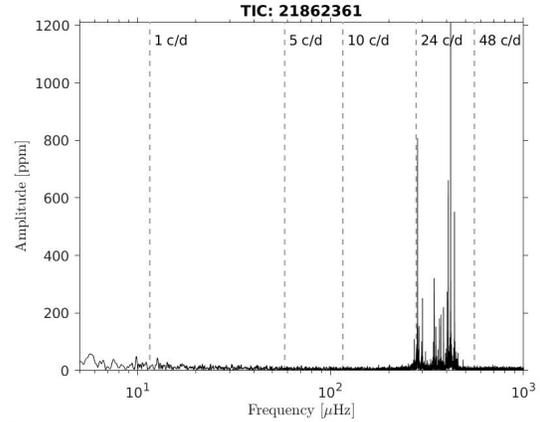

**Figure 2a:** Amplitude spectrum for HD 155316, shown to be a δ Sct variable from analysis of two sectors of TESS data. This spectrum shows 93 frequencies with S/N > 4.

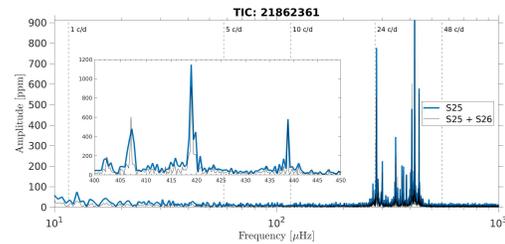

**Figure 2b:** Another view of amplitude spectrum of HD 155316. Using two vs. one sector of observations, the S/N ratio is increased by sqrt(2) (see low frequencies in inset), and the frequency resolution is improved from 0.46 μHz to 0.22 μHz.

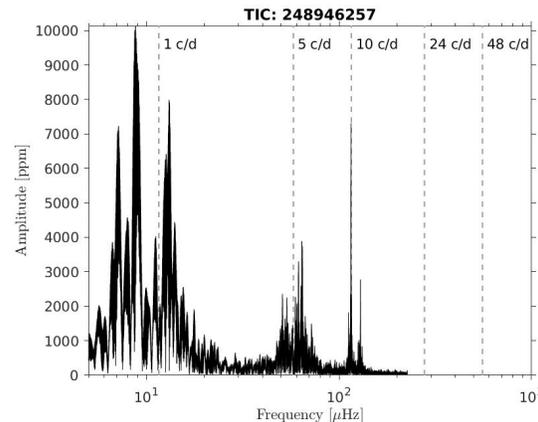

**Figure 3:** Amplitude spectrum for HD 8251, shown to be a δ Sct / γ Dor hybrid candidate from analysis of two sectors of TESS data. This spectrum shows 67 frequencies with S/N > 4.

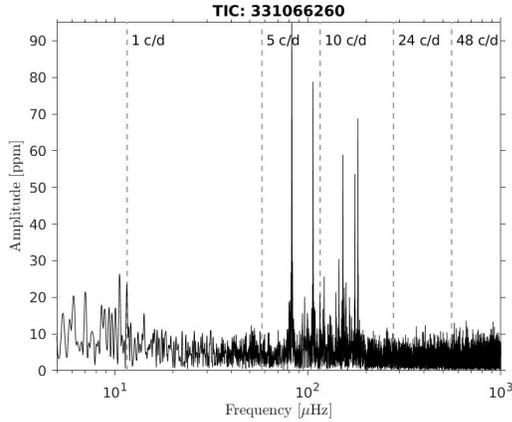

**Figure 4:** Amplitude spectrum for HD 211643, shown to be a δ Sct star from analysis of two sectors of TESS data. This spectrum shows 45 frequencies with S/N > 4. However, See section 4 re. possible contamination in the 2-min light curve data from a nearby star.

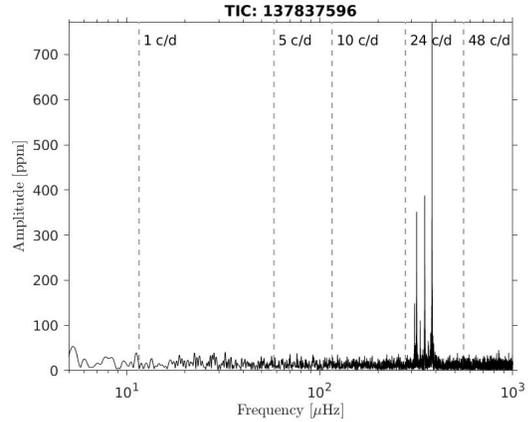

**Figure 7:** Amplitude spectrum for HD 212144, shown to be a δ Sct star from analysis of one sector of TESS data. This spectrum shows 10 frequencies with S/N > 4.

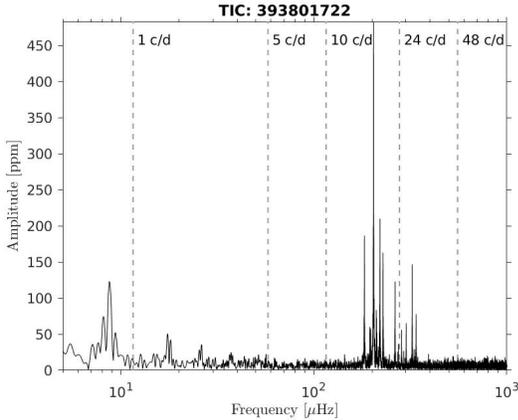

**Figure 5:** Amplitude spectrum for HD 108449, shown to be a δ Sct / γ Dor hybrid candidate from analysis of one sector of TESS data. This spectrum shows 34 frequencies with S/N > 4.

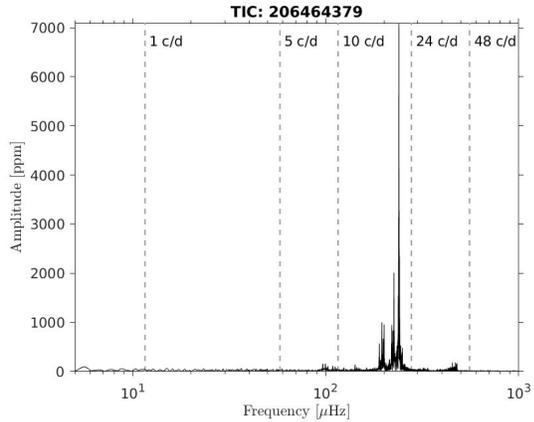

**Figure 8:** Amplitude spectrum for HD 209475, shown to be a δ Sct star from analysis of one sector of TESS data. This spectrum shows over 200 frequencies with S/N > 4.

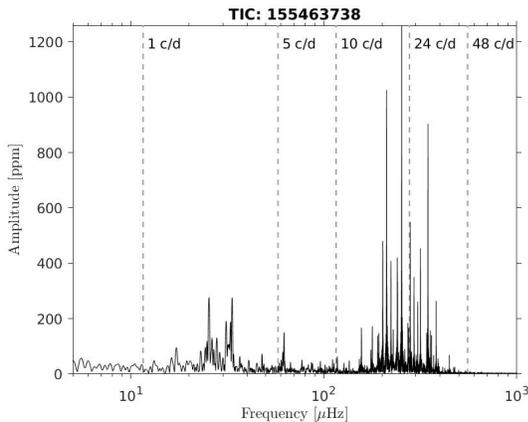

**Figure 6:** Amplitude spectrum for HD 50635, shown to be a δ Sct / γ Dor hybrid candidate from analysis of one sector of TESS data. This spectrum shows over 200 frequencies with S/N > 11.

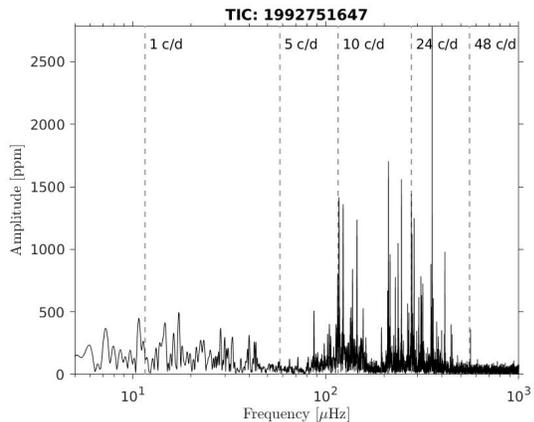

**Figure 9:** Amplitude spectrum for HD 196166, shown to be a δ Sct / γ Dor hybrid candidate from analysis of one sector of TESS data. This spectrum shows 156 frequencies with S/N > 4.

## 4. Contamination

One of the questions raised by these results is whether the pulsations are intrinsic to the star or a result of background or nearby star contamination. Some of the fields are relatively crowded, and the TESS pixel scale is 21 arc sec. Tables 1 and 2 list the TESS Input Catalog contamination ratios for each star when available; those above 0.01 are highlighted in bold. One hybrid candidate, HD 196166 (Fig. 9), as well as two δ Sct stars HD 211643 (Fig. 4) and HD 212144 (Fig. 7) have contamination ratios exceeding 0.01. The δ Sct frequencies for HD 211643 seem particularly suspect as their amplitudes are so low.

We used the lightkurve python tools introduced at the American Association of Variable Star observers Data Mining Workshop led by Colin Littlefield in November 2020 to examine the target pixel data. These tools allow one to create light curves with varying subsets pixels other those used to construct the 2-minute cadence light curve product in the TESS pipeline. For HD 196166 and HD212144, we cannot rule out a signal from a possible background or foreground star aligned with the brightest central pixels. The pulsation variations decrease or not present in the light curves generated using only pixels farther away from the central pixels.

The target pixel data for HD 211643 (Fig. 10a) shows a bright object above and to the right of the central star. We selected only four pixels centered on this object to generate the light curve. A frequency analysis of this light curve using AAVSO VStar software (Fig. 10b) shows an amplitude spectrum very similar to that of Fig. 4. Using the SIMBAD tool, we find the bright star about 85 arc sec (4 TESS pixel widths) away (Fig. 10c), and identify it as HD 211604, an A5 star with V mag 8.08, somewhat fainter than HD 211643 with V mag 7.13. HD 211604 is also not listed in SIMBAD as a known δ Sct variable.

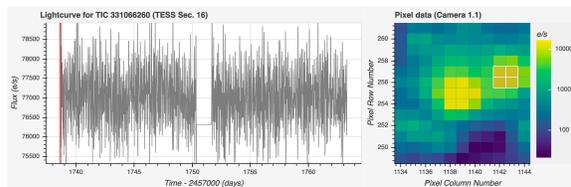

**Figure 10a:** Light curve generated from target pixel files for four pixels selected above and to the right of the target pixels for HD 211643.

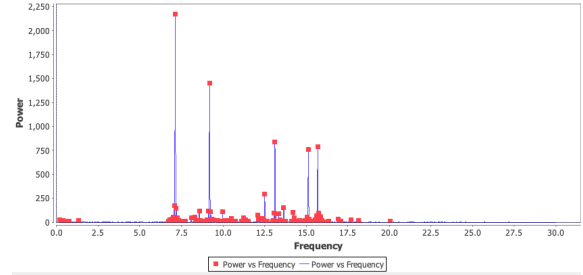

**Figure 10b:** Amplitude spectrum generated from light curve of selected pixels in Fig. 10a, showing δ Sct frequencies similar to those of Fig. 4.

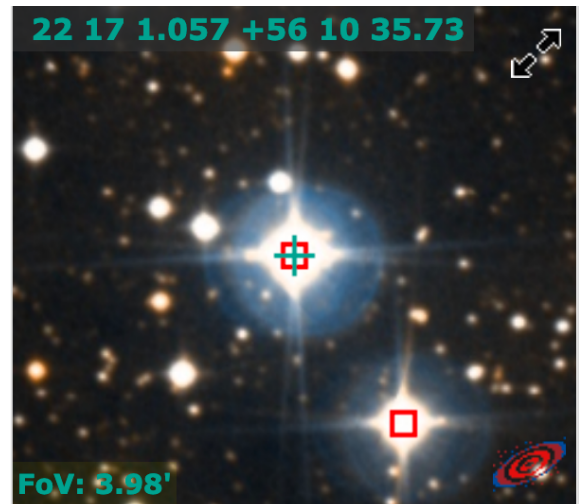

**Figure 10c:** SIMBAD search around HD 211643 shows at 85 arc sec separation in the lower right HD 211604, an A5 star with mag V=8.08 that is likely the source of the δ Sct pulsations. Note that the orientation of this figure differs from that of the TESS target pixel data in Fig. 10a.

## 5. HD 43682: Eclipse or planet transit?

HD 43682 (TIC 437984134) observed in Sector 33 shows a dip in its light curve with duration about 4.8 hours that might be caused by an eclipse or a planet transit. While the contamination ratio for this object is high, there are no nearby bright objects as was the case for HD 211643. We again used the lightkurve software to examine the target pixel file for this star (Fig. 11). The feature shows up only in the light curve of the brightest pixels, and disappears when only the surrounding pixels are selected. Therefore, the feature may be intrinsic to the star. The light curve dip only appears once during the 27-day time series.

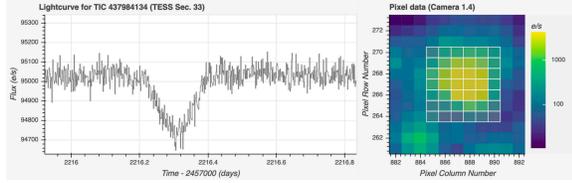

**Figure 11:** Zoom-in on TESS light curve for HD 43682 showing a dip of duration 4.8 hours. This feature does not show up on light curves generated using only nearby pixels, omitting the brightest pixels.

# 6. Locations of the Am stars on the H-R Diagram

Table 1 summarizes the properties of the 32 stars from the Catanzaro et al. (2019) sample. This table includes columns for the effective temperature and luminosity from the TESS Input Catalog version 8.1 (Stassun et al. 2019). This catalog merges data from a number of photometric and spectroscopic catalogs and the Gaia Data Release (DR2, Gaia Collaboration, 2018) parallaxes, on which these temperatures and luminosities are based. The table also includes the effective temperatures and luminosities derived by Catanzaro et al. (2019), also making use of Gaia DR2 parallaxes, but basing the effective temperatures and the flux derivations on their own high resolution spectro-polarimetry and spectral synthesis methods.

Figure 12 shows log $L/L_{sun}$ vs. $T_{eff}$ using the Catanzaro et al. (2019) $T_{eff}$ and luminosity determinations. Also shown are the theoretical red and blue edges of the δ Sct (Breger and Pamyatnykh 1998) and γ Dor (Warner et al. 2003) instability strips (IS). The two δ Sct stars fall within the IS, and one of the two hybrid stars falls within the region where the instability strips overlap.

Figure 13 shows the H-R diagram using the TESS Input Catalog version 8.1 $T_{eff}$ and luminosity. Using this data set, one of the hybrid stars moves to significantly cooler temperature, and now lies within the region where the instability strips overlap.

Figure 14 shows $T_{eff}$ derived from the TESS Input Catalog v8.1 vs. $T_{eff}$ from Catanzaro et al. (2019) spectroscopy for each star. The TESS Input Catalog v8.1 $T_{eff}$ values are slightly hotter than the Catanzaro et al. values for the hottest stars in the sample.

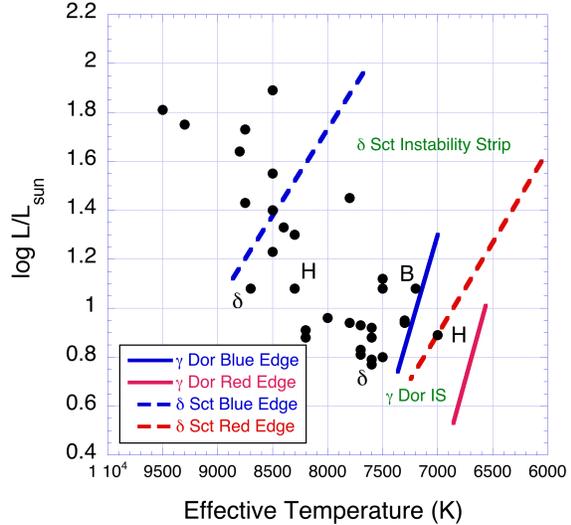

**Figure 12:** Location in H-R diagram of 32 Am stars observed by TESS (black dots), based on $T_{eff}$ and luminosity (L) determinations by Catanzaro et al. (2019). Also shown are the theoretical blue edge and observational red edge of the δ Sct instability strip (dashed lines) from Breger and Pamyatnykh (1998) and the theoretical red and blue edges of the γ Dor instability strip (solid lines) from Warner et al. (2003). The stars labeled are δ = δ Sct, H = hybrid candidate, and B = eclipsing binary, as determined from the TESS data.

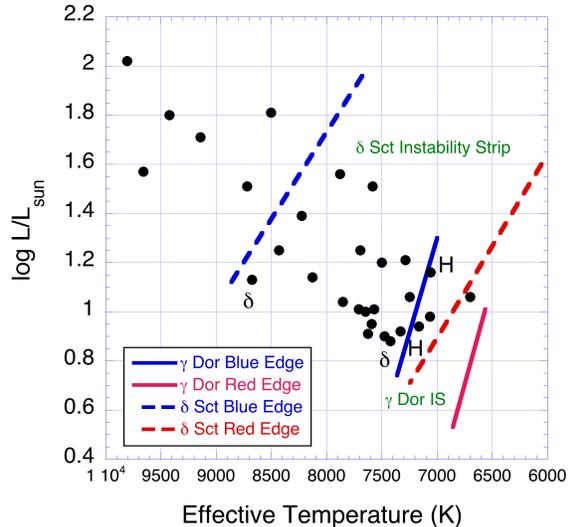

**Figure 13:** Same as Fig. 12, except using $T_{eff}$ and luminosity from TESS Input Catalog v8.1 (Stassun et al. 2019). Note that one of the hybrids has a significantly lower temperature than determined by Catanzaro et al. (2019), so that it is now located in the region where the instability strips overlap.

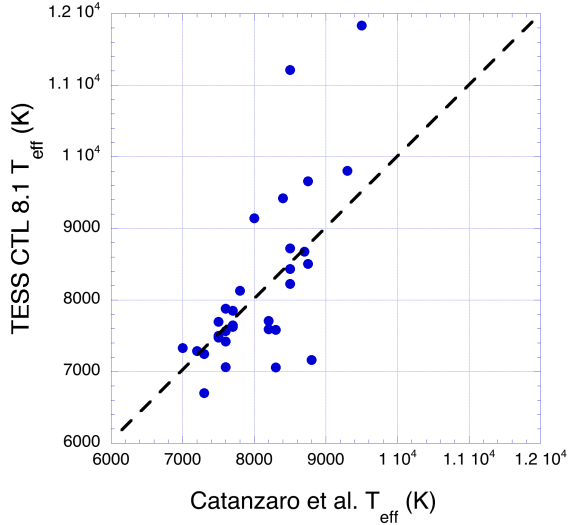

**Figure 14:** Comparison of effective temperatures determined by Catanzaro et al. (2019) and as listed in the TESS Input Catalog v8.1 for each star of Table 1. The black dashed line is the trend for perfect correlation.

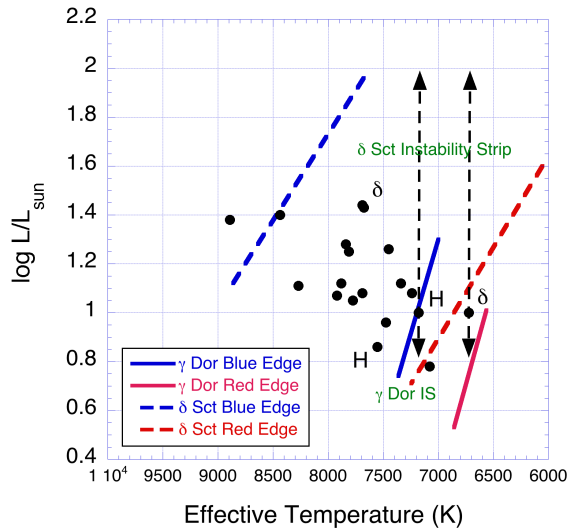

**Figure 15:** Location of 20 stars with $T_{eff}$ < 10,000 K in sample of chemically peculiar stars selected from Paunzen et al. (2013). The $T_{eff}$ and luminosity (L) as found in TESS Input Catalog v8.1 (Stassun et al. 2019). For two of the stars, the luminosity was not available, so these are plotted at log $L/L_{sun}$ = 1 with dashed vertical arrows. The two δ Sct stars and two hybrid candidates are indicated.

Figure 15 shows the stars from the Paunzen et al. (2013) sample observed by TESS plotted on the H-R diagram. Two of the stars did not have a luminosity in the TESS Input Catalog CTL v8.1, so they are plotted at log $L/L_{sun}$ = 1.0 with large dashed vertical arrows. All of these stars are in or near their expected pulsation instability regions.

## 7. Conclusions

Of the 32 Am stars in Catanzaro et al. (2019) sample observed by TESS so far, we find two δ Sct stars and two δ Sct/γ Dor hybrid candidates. However, in the TESS 2-min cadence light curve for HD 211643, the δ Sct pulsations likely originate from contamination by a nearby A-type star in the TESS field that may or may not be an Am star. Of the 23 stars in the Paunzen et al. (2013) sample observed so far, we find two additional δ Sct and two additional δ Sct/γ Dor hybrid candidates. As discussed in the introduction, Am stars are not necessarily expected to show δ Sct pulsations because the diffusive processes responsible for the peculiar element abundances also should deplete helium from the pulsation driving region and inhibit pulsations. The frequencies observed in these stars, along with fundamental properties such as effective temperature, luminosity, and element abundances, will provide useful constraints for asteroseismic modeling to understand the origin of the pulsations.

Much more work is needed to understand the causes of photometric variability seen in many of the stars that do not show δ Sct pulsations, and to confirm γ Dor g-mode pulsations in the hybrids. The stars in these samples are bright and amenable to ground-based photometric and spectroscopic observations, including observations by amateur observers. Some possibilities include time-series spectroscopy and multicolor photometry, useful for mode identifications and distinguishing between causes of variability, and to determine whether amplitudes or frequency content change over time.

## 8. Acknowledgements

We are grateful for data from the TESS Guest Investigator programs. J.G. acknowledges a Los Alamos National Laboratory Center for Space and Earth Sciences grant CSES XWPB ARR0GZIK and support from LANL, managed by Triad National Security, LLC for the U.S. DOE's NNSA, Contract #89233218CNA000001.

This research has made use of the SIMBAD database, operated at CDS, Strasbourg, France, the Mikulski Archive for Space Telescopes (MAST), and the TESS field of view search tool of the TESS Asteroseismic Science Consortium. J.G. also thanks the AAVSO and Colin Littlefield for training in using TESS data at the Data Mining Workshop held online in November 2021. J.G. thanks the Society for Astronomical Sciences for the opportunity to present this paper.

**Table 1: Summary of TESS results for 32 of the 62 stars in the Catanzaro et al. (2019) sample. Contamination ratios > 0.01 are highlighted in bold.**

| HD # | TIC ID | TESS Mag. | TESS CTL 8.1 $T_{eff}$ (K) | Catanzaro $T_{eff}$ (K) | TESS CTL 8.1 log $L/L_{sun}$ | Catanzaro log $L/L_{sun}$ | Contamin. Ratio | # Sectors Obs. | Result |
|---|---|---|---|---|---|---|---|---|---|
| 267 | 176299362 | 8.037 | 7708 | 8200 | 1.01 | 0.91 | 0.00003 | 1 | |
| 8251 | 248946257 | 8.173 | 7058 | 8300 | 1.16 | 1.08 | 0.00019 | 2 | **Hybrid cand.** |
| 10088 | 151056397 | 7.618 | 7647 | 7700 | 1.00 | 0.81 | 0.00498 | 1 | |
| 14825 | 445666347 | 7.729 | 8432 | 8500 | 1.25 | 1.23 | **1.27093** | 1 | |
| 99620 | 310922962 | 7.529 | 7853 | 7700 | 1.04 | 0.93 | **0.06044** | 3 | |
| 108449 | 393801722 | 8.036 | 7330 | 7000 | 0.92 | 0.89 | 0.00025 | 1 | **Hybrid cand.** |
| 117624 | 95526455 | 8.113 | 7501 | 7500 | 1.20 | 1.08 | 0.00118 | 1 | |
| 127263 | 158035384 | 7.928 | 7592 | 8200 | 0.95 | 0.88 | 0.00017 | 2 | |
| 132295 | 298247573 | 8.805 | 6701 | 7300 | 1.06 | 0.95 | 0.00121 | 2 | |
| 139939 | 137031010 | 7.305 | N/A | 7800 | N/A | 0.94 | N/A | 1 | |
| 143914 | 313423751 | 7.758 | 8128 | 8000 | 1.14 | 0.96 | 0.00010 | 4 | |
| 149650 | 198210258 | 5.913 | 9141 | 8800 | 1.71 | 1.64 | 0.00041 | 13 | |
| 149748 | 198212148 | 7.063 | 7165 | 7600 | 0.94 | 0.77 | 0.00026 | 12 | |
| 154226 | 21673730 | 7.752 | 7879 | 7800 | 1.56 | 1.45 | **0.01308** | 1 | |
| 155316 | 21862361 | 7.897 | 7423 | 7600 | 0.88 | 0.79 | 0.00260 | 2 | **delta Sct** |
| 164394 | 329343467 | 7.308 | 7625 | 7700 | 0.91 | 0.83 | **0.14943** | 6 | |
| 166894 | 76038913 | 7.694 | 8504 | 8750 | 1.81 | 1.73 | 0.00128 | 1 | |
| 167828 | 158178074 | 7.168 | 9658 | 8750 | 1.57 | 1.43 | 0.00385 | 1 | |
| 168796 | 157623505 | 7.780 | 8721 | 8500 | 1.51 | 1.40 | 0.00882 | 1 | |
| 169885 | 21162252 | 6.236 | 8226 | 8500 | 1.39 | 1.55 | 0.00179 | 7 | |
| 171363 | 289375356 | 7.820 | 7569 | 7600 | 1.01 | 0.92 | 0.00187 | 1 | |
| 180347 | 298969563 | 8.181 | 7063 | 7600 | 0.98 | 0.88 | 0.00538 | 3 | |
| 184903 | 267813367 | 7.813 | 9806 | 9300 | 2.02 | 1.75 | 0.00642 | 9 | |
| 188103 | 209945493 | 8.070 | 11836 | 9500 | 2.06 | 1.81 | **0.05970** | 1 | |
| 188854 | 268380299 | 7.317 | 7287 | 7200 | 1.21 | 1.08 | 0.00333 | 2 | Eclipsing binary |
| 189574 | 172758318 | 7.530 | 7476 | 7500 | 0.90 | 0.80 | **0.02421** | 2 | |
| 190145 | 366490533 | 7.370 | 7695 | 7500 | 1.25 | 1.12 | 0.00398 | 13 | |
| 192662 | 380794470 | 8.666 | 9424 | 8400 | 1.80 | 1.33 | **0.04673** | 2 | |
| 202431 | 388078603 | 7.203 | 7248 | 7300 | 1.06 | 0.94 | 0.00593 | 5 | |
| 210433 | 328745733 | 7.229 | 11213 | 8500 | 2.02 | 1.89 | N/A | 3 | |
| 211643 | 331066260 | 7.068 | 8675 | 8700 | 1.13 | 1.08 | **0.06081** | 2 | **delta Sct** |
| 215327 | 128920773 | 8.000 | 7583 | 8300 | 1.51 | 1.30 | 0.00959 | 1 | |

**Table 2: Summary of TESS results for 23 stars of the Paunzen et al. (2013) sample. Contamination ratios > 0.01 are highlighted in bold.**

| HD # | TIC ID | TESS Mag. | TESS CTL 8.1 $T_{eff}$ (K) | TESS CTL 8.1 log $L/L_{sun}$ | Contamin. Ratio | # Sectors Obs. | Result |
|---|---|---|---|---|---|---|---|
| 3644 | 212982053 | 8.692 | 7881 | 1.12 | 0.00194 | 1 | |
| 4570 | 333557388 | 9.026 | 7450 | 1.26 | **0.01778** | 1 | |
| 19963 | 365425158 | 7.663 | 7476 | 0.96 | 0.00002 | 1 | |
| 21933 | 416713269 | 5.838 | 11435 | 2.01 | 0.00143 | 1 | |
| 28355 | 245794190 | 4.858 | 7812 | 1.25 | 0.00104 | 1 | |
| 28546 | 245819988 | 5.273 | 7775 | 1.05 | 0.00026 | 1 | |
| 29479 | 245911904 | 5.000 | 8436 | 1.40 | 0.00691 | 1 | |
| 29589 | 245936817 | 5.559 | 14089 | N/A | N/A | 1 | |
| 31373 | 436723855 | 5.873 | 13456 | N/A | N/A | 1 | |
| 43682 | 437984134 | 8.165 | 7340 | 1.12 | **0.07555** | 1 | |
| 50635 | 155463738 | 4.407 | 7178 | N/A | N/A | 1 | **Hybrid cand.** |
| 68099 | 334411408 | 6.170 | 12804 | N/A | N/A | 1 | |
| 79066 | 286055622 | 6.023 | 7079 | 0.78 | 0.00225 | 1 | |
| 196166 | 1992751647 | 9.555 | 7553 | 0.86 | **3.10795** | 1 | **Hybrid cand.** |
| 197889 | 1992945158 | 7.251 | 7919 | 1.07 | **1.47153** | 1 | |
| 200052 | 29671013 | 6.024 | 8892 | 1.38 | 0.00251 | 1 | |
| 202152 | 1992873233 | 8.311 | 7689 | 1.08 | **2.12545** | 1 | |
| 203313 | 302334615 | 9.160 | 7839 | 1.28 | 0.00739 | 1 | |
| 203653 | 300272058 | 7.819 | 7690 | 1.44 | 0.00017 | 1 | |
| 209475 | 206464379 | 7.644 | 7675 | 1.43 | 0.00380 | 1 | **delta Scuti** |
| 212144 | 137837596 | 9.850 | 6723 | N/A | **0.01085** | 1 | **delta Scuti** |
| 212797 | 69750857 | 8.355 | 7239 | 1.08 | 0.00506 | 1 | |
| 215875 | 146165482 | 8.132 | 8270 | 1.11 | 0.00002 | 1 | |